\documentclass{ppig}
\usepackage{epsfig}
\usepackage{booktabs}
\usepackage{ucs}
\usepackage[utf8x]{inputenc}
\usepackage[dvipsnames]{xcolor}
\usepackage{xspace}
\usepackage{enumitem}
\setlist[itemize]{nosep}
\setlist[enumerate]{nosep}

\newcommand{\BingChat}{Bing Chat\xspace}

\title{Participatory prompting: a user-centric research method for eliciting AI assistance opportunities in knowledge workflows}

\author{Advait Sarkar \\
  Microsoft Research \\
  University of Cambridge \\
  University College London \\
  advait@microsoft.com \\
  \And
  Ian Drosos \\
  Microsoft Research \\
  t-iandrosos@microsoft.com \\
  \And
  Rob Deline \\
  Microsoft Research \\
  rob.deline@microsoft.com \\
  \AND
  Andrew D. Gordon \\
  Microsoft Research \\
  University of Edinburgh \\
  adg@microsoft.com \\
  \And
  Carina Negreanu \\
  Microsoft Research \\
  cnegreanu@microsoft.com \\
  \And
  Sean Rintel \\
  Microsoft Research \\
  serintel@microsoft.com \\
  \AND
  Jack Williams \\
  Microsoft Research \\
  jack.williams@microsoft.com \\
  \And
  Ben Zorn \\
  Microsoft Research \\
  ben.zorn@microsoft.com \\
}
\date{}

\begin{document}
\maketitle
\thispagestyle{empty} % UNCOMMENT BEFORE SUBMISSION

\begin{abstract}
Generative AI, such as image generation models and large language models, stands to provide tremendous value to end-user programmers in creative and knowledge workflows. Current research methods struggle to engage end-users in a realistic conversation that balances the actually existing capabilities of generative AI with the open-ended nature of user workflows and the many opportunities for the application of this technology. In this work-in-progress paper, we introduce participatory prompting, a method for eliciting opportunities for generative AI in end-user workflows. The participatory prompting method combines a contextual inquiry and a researcher-mediated interaction with a generative model, which helps study participants interact with a generative model without having to develop prompting strategies of their own. We discuss the ongoing development of a study whose aim will be to identify end-user programming opportunities for generative AI in data analysis workflows.
\end{abstract}
% \benz{What defines "realistic"} \ian{addressing the open-ended nature of actual user workflows, rather than in-lab style structured tasks}

% \benz{Top-level comments: inconsistent capitalization of Bing "chat" or "Chat".} \ian{good point, I will make a macro for this}
\section{Introduction and motivation}
% PPIG 2023 call for papers: \url{https://www.ppig.org/workshops/2023-annual-workshop/}. Abstract deadline May 15, paper deadline June 1.

Generative AI presents many opportunities for assistance and automation for end-users and end-user programmers. Our research team is interested in exploring how Large Language Model (LLM) assistance can be used in data-driven sensemaking \cite{russell1993cost,pirolli2005sensemaking} in spreadsheets, to identify key areas of strength and weakness in LLM assistance and identify opportunities for LLM assistance that address specific parts of the overall workflow. End-user data analysis workflows are complex and range over many steps, including problem conceptualization, identifying relevant datasets, data cleaning and structuring, developing an analysis strategy, learning how to use relevant features, open-ended exploration, and presenting results.

% 1.	Problem conceptualization, decomposition, identifying parts of the problem that could be tackled in a spreadsheet
% 2.	Identifying relevant datasets
% 3.	Figuring out how to clean and structure data
% 4.	Developing an analytical strategy, involving application of multiple features in multiple steps
% 5.	Learning how to use relevant features
% 6.	Exploration of alternative analyses
% 7.	Presenting and communicating results

The effect of generative AI on knowledge work has been described as a shift \emph{``from material production to critical integration''} \cite{sarkar2023critical}. Critical integration consists of \emph{``deciding where in the workflow to use the productive power of AI, how to program it correctly [...], and how to process its output in order to incorporate it''}. Sarkar builds on the theory of double-loop learning in organizations \cite{argyris1977double}, observing that there is both an inner-loop aspect to applying AI in knowledge workflows (incorporating AI assistance in various steps of existing workflows) as well as an outer-loop aspect (reconfiguring knowledge workflows to take better advantage of AI, and developing new ones which are only possible with AI). 

For example, in data-driven sensemaking, critical integration in the inner loop might consist of finding applications for AI in data visualization, or data cleaning. Critical integration in the outer loop might consist of applying AI towards identifying a suitable analysis strategy or automating large portions of the sensemaking workflow (e.g., in the spirit of the ``automatic statistician'' \cite{steinruecken2019automatic}) and developing new tools for human overseers focusing on auditing and quality control.

A key question for researchers and designers at this point in time is how to study the needs of users involved in such workflows. We are concerned with the first phase of the design ``double diamond''; we first need to design the right thing, and only later can we attend to getting the design of the thing right \cite{buxton2010sketching}. As generative AI technology is new and continuously evolving, its use in society is limited and uneven. It may not be possible, for example, to simply observe participants working with generative AI, or interview them about their work practices with generative AI, if generative AI is not widely adopted within their workflows (which is the case for the vast majority of knowledge work at the time of writing). For example, code completion in code editors for professional software developers has been an early commercialization of generative AI, which gives researchers a wide pool of experienced users with mature behaviours to study \cite{sarkar2022programmingai}. Other work has discussed code assistants for data analysis within computational notebooks \cite{McNutt2023DesignOfAICodeAssistants}. Unfortunately, for our end-user scenario of interest (data analysis workflows in spreadsheets), this is not yet the case. 
% \benz{Also - generative AI in GitHub Copilot follows the previous Intellisense tech, with which there was much experience.  Also might want to cite related work from our MSR colleagues in CHI this year: On the Design of AI-powered Code Assistants for Notebooks https://dl.acm.org/doi/abs/10.1145/3544548.3580940}

Furthermore, it is not ideal for researchers to develop high-fidelity experiences for generative AI as a way of testing its applicability to different workflow. It is time-consuming and expensive. Moreover, it is also limiting; out of the wide variety of potential interventions at the inner and outer loops, only a very small number can be feasibly explored using a functional prototype.

The traditional solution to this has been to use lower-fidelity methods such as Wizard-of-Oz \cite{gould1983composing,landauer1986psychology}, paper prototyping \cite{snyder2003paper} and champagne prototyping \cite{blackwell2004champagne}, which allow researchers to rapidly simulate a wide variety of user experiences with significantly lower engineering costs, while also enabling interaction with experiences that may be extremely challenging or impossible to build due to technical limitations. However, these methods have limitations as well; for a Wizard-of-Oz study to have direct implications for design, the Wizard protocol must correspond to the actually existing capabilities of the system(s) that are eventually built. 

In particular, the mythologizing of AI's capabilities by the media, academia, and industry has led to a warped public conception of what AI can do and how it works \cite{sarkar2022explainable}. Thus Siddharth et al. \cite{siddarth2021ai} urge us to focus not on this collective mirage of what AI might be, but on ``actually existing AI (AEAI)''. There is a real risk that participant responses in low-fidelity studies will draw from their own biased and inflated expectations of AI capabilities to fill in the ``gaps'' left by the incomplete nature of the prototype. A poorly designed Wizard protocol, which allows too much improvisational deviation from a script, can exacerbate this. This problem is even greater in generic ``need-finding'' interviews where no prototypes are used.

There is thus a need for a research method that combines the advantages of low-fidelity methods such as Wizard-of-Oz, and rapidly exploring a wide range of potential interactions at both the inner and outer loops of a knowledge workflow, while still grounding conversations with participants in the capabilities of actually existing AI. In response, we have been developing a method called \textbf{participatory prompting}.

The participatory prompting method takes the form of a researcher-mediated interaction between a study participant and a working generative AI system. During the session, the researcher guides the participant through a workflow, seeking to test the potential applications for AI at each step. The researcher plays multiple roles in facilitating this interaction. Most importantly, they restructure user requests according to pre-identified prompting strategies, and help the user continue the interaction and recover from errors. The semi-structured interview is grounded in a specific real problem of interest to the user, drawing on principles of contextual inquiry \cite{raven1996using}. The name of the method is inspired by participatory design \cite{spinuzzi2005methodology}, and we hope that in the spirit of participatory design, the method of participatory prompting contributes to the design of AI systems that empower and enfranchise users with their involvement from the outset. The next section describes the method in detail.

% Data-driven sensemaking \cite{} is one such workflow.

% \section{Related work}
% Todo: explain
% Participatory design
% Wizard of Oz
% Contextual inquiry
% Champagne prototyping
% Semi-structured interviews

\section{The participatory prompting method}

\subsection{Materials required}

\paragraph{Choice of system.} The participatory prompting method uses a real, functional generative AI system as representative of the functionality of generative AI in general. It is therefore important to choose the system carefully and consider multiple alternatives for their suitability to the particular study.

We compared the following four systems for their suitability for use in our study: OpenAI playground, OpenAI ChatGPT, Google Bard, and Microsoft \BingChat. We compared them by entering some example queries that a user might have into each system, and attempting to elicit guidance and multiple stages of the data analysis process, as we were intending to do during the study. We then discussed and evaluated the comparative quality of the responses and how a participant might react to each response, with a view to choosing the system which would help produce the most insightful interactions during the study.
% \benz{Was \BingChat in Creative mode?  That's when you get GPT-4.} \ian{for the initial testing we tried out different modes of Bing, but yeah eventually settled on creative mode. I don't know if it is public that creative is the gpt4 mode and the others aren't though?}

It is worth noting that of the four systems we tested, the latter three (ChatGPT, Bard, and \BingChat) are consumer-facing products: they are built upon one or more large language models and consist of UI elements and other modules and heuristics which come together to create a coherent experience for the non-expert consumer. They can be considered to be significantly ``opinionated'' in a number of ways. One obvious user-facing manifestation of this is in the so-called ``guardrails'' which kick in whenever the conversation topic approaches an area deemed inappropriate by the system designers (such as violent or sexual content). Another example of opinionation is the turn limits imposed by \BingChat: at the time of writing, a conversation with \BingChat cannot exceed 15 turns (after 15 turns, the conversation is erased, and a new conversation is started).

In contrast, the OpenAI playground is intended for developers to interactively test different models, and as such allows for the choice between multiple individual LLMs, and control over parameters such as temperature (most of our testing on the OpenAI playground was using the default temperature of \texttt{0.7} and the \texttt{text-davinci-003} model, a GPT-3 model which was the state of the art at the time of testing). While there are still some heuristics and guardrails in place, using the OpenAI playground is much closer to getting the ``raw'' output of a language model.

For many studies, using a highly opinionated experience may not be ideal; the heuristics and modules used in these systems are proprietary, and researcher control and visibility into parameters such as temperature is poor. For our purposes, however, this was not a dealbreaker.

For our study we have chosen to use \BingChat for the primary reason that it is the only system (at the time of writing) that is designed to seek and include information from the Web as part of its responses. In our testing, we found that for many steps of the data analysis journey (ideating potential analysis paths, identifying relevant datasets, learning to use relevant features), the ability to report information from the Web resulted in much better and more actionable suggestions for users.
% \benz{Yes, access to Search is a big differentiator although Bard should also have this, no? Another aspect of \BingChat is that there are different models that you possibly get - thus it's a bit harder to say exactly what the quality is because it changes.} \ian{on bard: at the time Bard did not cite sources (and refused to when asked), not sure on the current status though. On model changing, as above I'm not sure it is public and was unable to find reports on it after a websearch}

Due to the complexity of deploying these systems at scale, during our comparative evaluation we noticed many outages, where the system was overloaded and did not respond to queries and/or displayed an error message. For a user study to go smoothly, a system that is stable and consistent is key. While we did not quantify the outages we experienced, our informal assessments of a particular system's reliability and uptime did influence our final decision.

% Choosing an LLM instance to use (e.g., OpenAI playground, which version of GPT3). Question: how can we get access to an LLM instance that is stable and consistent enough for a user study. OpenAI playground has periodic outages, for example, which we don’t want during a study. 

%  \BingChat’s clear advantage is that it can search the web. However, it is opinionated (has guardrails, turn limit) and we have limited access to internal settings (temperature etc.).

\paragraph{Prompt strategies.} Identifying performant and consistent strategies for prompting LLMs is a well-documented challenge. In consumer-facing products, the user query is rarely sent directly to an LLM; instead it is processed and augmented with additional instructions and prompts that have been determined by the system developers. 

Thus, when non-experts directly interact with a ``raw'' LLM (e.g., via tools such as OpenAI playground), or with a generic chat application that is not tuned towards particular knowledge workflows, they may not be able to develop suitable prompting strategies to elicit good performance from the model. This is a key reason that our method involves researcher-mediated interaction, and why we do not simply study how end-users interact directly with the model.

A key strength of the participatory prompting method is that researchers familiar with the design of prompting strategies can prepare these ahead of time. 

For our study, a group of researchers collaboratively experimented with different prompting strategies with \BingChat over a period of several weeks, documenting screenshots of their interactions with \BingChat and successful prompts in a shared document. A provisional list of prompting strategies which we developed through this process is given in Appendix~\ref{apx:prompts}. For example, through this process we identified that \BingChat:
\begin{itemize}
    \item did not consistently use data sources from the web even if they were available, and we could bias it towards doing so by including a phrase in the prompt such as ``use an online data source'', ``based on publicly available information'', ``with data from the web'', and ``use information from the web''.
    \item did not consistently offer citations for sources, but could be biased to do so by including a phrase in the prompt such as ``prove your sources are real'' or ``cite your sources''.
    \item often provided multiple suggestions for types of data analysis the user could conduct, but the answers did not support an end-user's decision for what to do next. In this case we found that adding ``justify your answer'' or ``justify your criterion'' improved the actionability of the model's responses.
    \item is capable of rendering tables inline within the chat, which is very helpful for exploring ideas related to spreadsheet-based data analysis, but it does not consistently do so. We found that we could bias it towards generating tables by specifying ``with an example'', ``make an example spreadsheet'', or ``make an example table''.
\end{itemize}

Our method for identifying prompts is largely a pragmatic craft practice, based on trial-and-error and the intuitions of researchers. Due to the many sources of variability in LLM output, as well as variability between researchers' experience and the working examples they choose for testing different prompting strategies, our resulting prompts are subjective and difficult to reproduce. Another team, or the same team choosing different working examples, or a different model, may well have developed a different set of prompts, which will have significant downstream effects on the user study. Improving the consistency and systematicity of this step is a major challenge for user research with generative AI, as many libraries, toolkits, and even prompt marketplaces have been created to assist in this endeavour.
% \benz{Mention that libraries and toolkits for this are a hot topic right now?}

% Another limitation is that, as a consumer-facing tool, \BingChat is likely to add its own post-processing and additional prompts to the input before sending it to the LLM. Researchers have no visibility or control over this. However, as a 

\paragraph{Demographics and generative AI experience.} Participants will complete a standard demographics questionnaire which includes questions about spreadsheet experience, formula experience, and programming experience \cite{sarkar2020spreadsheet}. In future participatory prompting studies, this can be replaced with another demographics questionnaire that gathers information relevant to those studies instead.

% \paragraph{Generative AI experience.} 

Based on the model of other questions in that questionnaire, we also developed a simple questionnaire item for assessing prior experience with generative AI, as follows: \emph{``Which of the following BEST describes your experience with generative AI tools such as ChatGPT, DALL-E, Stable Diffusion, Midjourney, Google Bard, \BingChat?''}

In response, participants choose from the following options:

\begin{enumerate}
    \item Never heard of them
    \item Heard of them but haven't tried any
    \item Casually tried one or more
    \item Occasionally use one or more
    \item Regularly use one or more
\end{enumerate}

As with other studies which use the aforementioned spreadsheet experience questionnaire, this item can be used in one of two ways: first, it can be used as part of the qualitative interpretation of participant interview data, to add context to their responses. Second, it can be used to group participants into rough categories of high and low prior experience (e.g., response levels 1-3 can be considered ``low'' experience and 4-5 can be considered ``high'' experience) for studying quantitative interactions between experience and any dependent variables gathered during the study (e.g., cognitive load \cite{hart1988development}).

Unlike spreadsheet experience or programming experience, the landscape of end-user experience with generative AI is shifting rapidly. The specific wording of this question and its response categories are thus likely to require periodic revision and updates.

\subsection{Main interview activity}
The main phase of the participatory prompting study takes the form of a semi-structured interview run concurrently with a researcher-mediated ``conversation'' between the participant and the model.

% A single turn is illustrated in Figure~\ref{fig:turn}.

The interview consists of a number of ``turns'' consisting of 5 steps. (1) A turn begins by the participant expressing a query (e.g., asking for assistance, posing a question, asking for clarification). (2) Next, the researcher takes the user query, modifies and augments it according to the previously identified prompting strategies and sends it to the model. (3) The participant reads the model's response. (4) Next, the researcher asks the participant to reflect on the response. (5) Finally, the researcher guides the participant in continuing the conversation and choosing the next query.

We wish to explore the possibility for LLM assistance in the following scenarios:
\begin{itemize}
    \item Problem conceptualization, decomposition, identifying parts of the problem that could be tackled in a spreadsheet
    \item Identifying relevant datasets
    \item Figuring out how to clean and structure data
    \item Developing an analytical strategy, involving applying multiple features in sequence
    \item Learning how to use relevant features
    \item Exploration of alternative analyses
    \item Presenting and communicating results
\end{itemize}

The problem chosen is ideally seeded by the participant's own problem domain. This can be elicited using a question such as: \emph{``Can you share an example of a decision you had to make recently? The decision should be reasonably complex, requiring an evaluation of multiple criteria or sources.''} 

If elicited ahead of the study (e.g., in a pre-study communication, or as part of the initial demographics questionnaire), researchers could prepare a spreadsheet and problem that is familiar to the participant’s own experience, or we can have the participant bring a shareable spreadsheet within their domain to work on. Alternatively, a suitable problem can be determined at the start of the interview. In practice we have found it is better to ask participants to think of such problems ahead of time, so that more time can be spent on the interactive portion of the interview.

The problems users bring can further be divided into the following types:
\begin{enumerate}
    \item A well-established spreadsheet workflow where the user is already using spreadsheets.
    \item An open-ended problem where the user has not tried to apply spreadsheets before.
\end{enumerate}

From a research perspective, both types of seed problem have advantages, as they correspond respectively to the inner and outer loop of the double loop of AI assistance opportunities. In our study context, we are not interested in one or the other type in particular, or in comparing between the two, so we will not aim to control the distribution of types. However, future studies may be interested mainly in inner loop opportunities, or outer loop opportunities, or a direct comparison between them. In such cases care must be taken to ensure the seed problems used with participants are either predominantly of the type of concern, or roughly evenly distributed between the types to facilitate comparison.

With a suitable seed problem, we walk through the participant's problem step by step, entering their requests into the LLM system (using pre-identified prompt strategies) and relaying their response back to the user. We find that it is useful for participants themselves to also view the screen on which the LLM interaction is taking place, as the study can progress faster when participants can read the output directly themselves.

% \benz{Is there an effort to know if the participant already tried using a model to solve the problem before participating in the study?  It might relate to understanding how the mediation was necessary for them. There's also the learning effect of the participant seeing how the mediated prompts are being phrased so they can use the same techniques in the future.} \ian{this is a good point for the pre-study interview for those who say they have chat experience and to ellicit in the post-study interview}

The user is then asked questions at each step such as:

\begin{itemize}
    \item Is this useful? Why or why not?
    \item Are you confused or surprised? Why or why not?
     \item Does it contain anything that is factually incorrect or misleading?
\end{itemize}

To advance the conversation to the next turn, the experimenter may prompt the participant with a question such as
\begin{itemize}
    \item Does this give you any further ideas?
    \item What would you like to know next, to continue your analysis?
    \item What information is missing?
\end{itemize}

Participants may also leverage suggested follow-up questions provided by the model as inspiration. However, since these suggestions may not adhere to experimenter prompting strategies, the suggestions may need intervention by the experimenter to align them.

% \benz{\BingChat already gives "follow up question" suggestions. 
%  Could these be used in the experiment?} \ian{since the participant can see the resulting output, we should include suggested questions as part of it. The issue with just straight up using the follow up recommendations is they might not include the prompting strategies we crafted and could run into issues we saw in the pilot}

\subsubsection{Post-activity interview}
% After the participant has walked through their problem with the LLM, we will ask follow up questions pertaining to design questions the team has, such as the placement of UI and the structure.

After the turn-taking phase of the study, the participant is interviewed and asked to reflect on the experience. For instance, they could be asked how such a tool would fit into their workflow, what features they feel would improve the experience, and what were the strengths and weaknesses of the new modes of working enabled with generative AI.

This phase may additionally involve eliciting participants' responses to mock-ups of potential interface designs in a design probe. Importantly, because the participant has just had the experience of interacting with an actually existing AI system, they are more likely to have an accurate understanding of what a different interface design might actually achieve for their workflow in terms of usability. The participant's grounding in actual AI capabilities improves the validity of research insights over simply interviewing participants about their response to mock-ups. 

If the participant reported that they had experience with generative AI, experimenters could elicit the participant to compare their previous workflows with what they experienced with participatory prompting. This might include the differences in solving similar or the same problems they saw in the study, or even understanding how the participant might change their own prompting strategies after the study.

The structure of and questions asked during the post-activity interview depends on the aims of the participatory prompting study. In our situation, we are interested in how generative AI can help non-expert end-users in data analysis workflows, particularly within spreadsheets, and so we selected our questions accordingly. Our full final script (incorporating revisions made after a pilot study detailed in Section~\ref{sec:pilot}) is given in Appendix~\ref{apx:script}.

\subsection{Pilot}
\label{sec:pilot}

The first version of this study protocol was piloted on a convenience sample of two participants who are familiar with spreadsheets and who use spreadsheets for their work. Each pilot took approximately 1 hour, as intended.

The pilots resulted in the following observations and adaptations:

It can be difficult for participants to settle on a suitable seed problem that is complex enough that it requires generative AI (as opposed to a traditional web search) to solve, but simple enough that the required context can be described to the system using a few sentences or a paragraph at most. We introduced more questions in the problem elicitation phase of the study that the experimenter could use to help the participant (e.g., \emph{``Can you share an example of a problem that required you to develop a new workflow?''}). However, our recommendation is that if possible, the participant should be asked to think of a suitable problem ahead of the scheduled study session, to maximize the time available to engage with the problem in the turn taking phase. We also noticed that participants were not familiar with some jargon in our questions (e.g., ``data-driven decision-making'') and we modified our questions to elaborate and clarify these terms.

We noticed that the participants were able to go through 5-6 turns in the time allotted. This may seem like a small number of turns, but it nonetheless produced a wide range of qualitative insights. The turn-taking phase can be elongated in future studies if this is felt to be necessary, study duration targets and participant fatigue notwithstanding.

The most time-consuming aspect of each turn is in the reflection step, where the participant is asked to reflect on the system's response, and the advancement step, where the researcher guides the participant to decide what to do next. This observation helped us decide on setting \BingChat to ``creative'' mode for the study. \BingChat has a single user-facing setting. The user can choose between creative mode (described by the \BingChat UI as ``original and imaginative''), balanced mode (``informative and friendly''), and precise mode (``concise and straightforward''). We initially used ``precise'' mode because we believed that it would be the least likely to hallucinate misinformation, and because generating short responses would allow the user to read through them more quickly and therefore enable more turns overall. Since the number of turns is largely dominated by the time spent on the reflection and advancement phases, the small time advantage gained in precise mode by having to read less text did not accumulate to allow an increased number of turns overall. Moreover, in practice we observed that ``creative'' mode was no more likely to generate hallucinations, and since it was far more verbose, often emitting several paragraphs in response, it improved participants' reflections (by giving them more to reflect on) and the ease with which a suitable next query was selected.

We noticed that if the model's response is completely generic or not useful, especially at an early stage of the conversation, our pilot participants were not motivated to continue the interaction. In response to this, we introduced a number of different advancement-oriented questions the researcher could use to help suggest a way forward, such as: \emph{``What would you change in your query to make this more useful? Would you ask this a different way?''}.

We noticed participants' preconceived notions about the system's capabilities were heavily influenced by their prior experience with search engines, and initially thought to use short queries of the kind used with search engines. This is unsurprising given \BingChat's positioning within a more traditional search interface. Previous studies have also noted that participants' use of generative AI systems is influenced by their experience with search engines \cite{liu2023wants,sarkar2022programmingai}. However, such short queries cannot adequately capture the user's context and intent. We introduced a guidance statement in the protocol whereby the experimenter explains that the generative AI system permits longer and more conversational interaction.

We also introduced a couple of strategies for the experimenter to gently suggest a way to continue the conversation, if the participant was having difficulties ideating a next step. These include the experimenter directly suggesting an action (e.g. \emph{``Let's see what happens if we try <some query>''}) but also suggesting an action as a baseline to help the participant conceive a contrastive alternative (e.g., \emph{``I propose to continue by <some query>, but what would you have done instead?''}). However, it is important that the experimenter's suggestions do not bias or significantly change the course of the interaction, and serve mainly to unblock the participant. Much as with regular interviewing, the ability to elicit rich responses from the participant without introducing bias depends on the skill of the interviewer. To help guard against such bias, we recommend in the protocol that these experimenter-led strategies should not be employed in consecutive turns (i.e., if the experimenter led the query in the previous turn, they should not do so in the current turn).

% ID pilot here: https://microsofteur-my.sharepoint.com/:w:/g/personal/t-iandrosos_microsoft_com/EVEH_YXtIDxPrglsbj7dEzQBSh0j0ZbmAjyVdiBnEEqA9A?e=PDg3RZ
% check slides for other notes as well

% Issue and result? AS -- the uncommented portion below is mainly a protocol issue, though as you noted the commented "RESULT" after that is a result.
There were issues with understanding participant expectations of model output, where even after working with the experimenter to craft a prompt, the participant did not know they would need to specifically request images. This led to needing to re-prompt the model to obtain the desired result. To prevent needing to do multiple prompts to obtain desired output, which can take up study time, experimenters should inquire on the expected results, including data types, from the participant to close this gap. Understanding what types of data could be useful for the participant, and explicitly requesting them in the prompt, is a necessary strategy.
% (AS -- IMO this continues the "issue") Understanding what types of data could be useful to the participant and explicitly requesting them in the prompt is necessary.
% RESULT: Participant wanted more multi-modal output (e.g., images of the locations she was interested in), when the model just returned text.

% RESULT? AS -- yes
% Participant was not sure the model considered the entire context of the prompt she gave it (e.g., the dates she would be visiting the city that she wanted open/close times for), even when this context was in the prompt.

% Issue and result?  better prompts to include previous outputs AS -- this feels mostly like a protocol issue (the revision being changing the prompting strategies) abut we should also discuss it in the limitations section limitation of the protocol (we can encourage the model but can't guarantee it).
Participants noticed that content the model had previously shown in the output could be missing in successive outputs. When the motivation of the participant was to build upon previous results, they wanted to make sure the data was consistent throughout the conversation with \BingChat. Therefore, prompts crafted for the study need to include or refer to previous output in an attempt to have the model consider this data for further prompts.

% RESULT? AS -- yes, a result
% After 3 prompts the participant noted she would start a spreadsheet to maintain the data she was receiving from the model, but thought it would be difficult to switch between the spreadsheet and further prompting she wanted to do. Getting the model to generate a table to help her visualized the future spreadsheet was helpful in this situation.

% Issue and result? better prompts to include the original context AS -- feels mostly like a result
% Model responded with content irrelevant to her context (e.g., ticket prices for children, seniors), which took the place of content she wanted to see in the table.

% Issue? AS -- since we don't have a good way to revise the protocol to get rid of this problem, we should discuss this in the limitations section.
% Another issue with the table the model presented was that it dropped data like links to photographs of the places she was interested in, and requested in a previous prompt, once the data being returned was converted into a table. This might be a limitation of the table view not being able to show links as the normal chat view can.

% Result but could impact on if we allow this? AS -- yes, we should discuss this as an issue, as a potential variation of the protocol -- allowing user to explore and verify is not explicitly part of the protocol but it could be.
One participant was suspicious of the data the model produced as a column in a table and wanted to verify this data by going to the websites the model referenced. This is a third workflow separate from prompting and spreadsheeting that requires a tangent into navigating to the source and verifying the data. This is a realistic strategy for users of chat based LLMs, but it is removed from prompting interactions. While this is not explicitly part of the protocol, we will allow participants the freedom to explore and verify the results returned by the model if desired.

In the post-activity interview, we noticed participants speculating on multiple occasions that \emph{``if it could do <some action>, that would be helpful''}. Since these types of questions can actually be put to the system to test whether it can do it, we amended the protocol to permit the researcher to spot-test such participant speculations and get feedback from the participant. We also introduced the following question to specifically elicit perceived barriers to sensemaking with AI assistance: \emph{``What barriers or frustrations did you have with this experience that prevented you from exploring the question to your satisfaction?''}

% \benz{The currrent availability of ChatGPT plugins specifically addresses the issue of "can the LLM do some action" by allowing users to let it.  Is the fact that this extensibility is already available something to be considered as part of the study in the future?} \ian{plugins are similar to any other tool study, I'm not sure how we could encompass the large amount of tool configurations against the current state of the art beyond an in-the-wild type study where participatory prompting does not apply}

Our full revised script after the pilot is given in Appendix~\ref{apx:script}.

\subsection{Effectiveness of protocol during pilot}
% AS: We aren't claiming that these are proper findings because it's only n=2 and we had to adapt the protocol as we went along so don't focus on those, but the point is that we're able to learn some actually useful stuff that (a) you couldn't learn just by WizOz because we used a real model (b) we were able to get these insights approximately as cheaply as WizOz because we developed prompts beforehand and didn't need to fine tune / build any new UI. Therefore, the protocol is a useful improvement over previous methods.
While we do not claim these are usable findings due to currently running a pilot of n=2 and the protocol was adapted live during these pilot runs, we believe there is evidence that this protocol was effective at revealing valuable insights from participants. These include:
\begin{itemize}
    % \item Participant blamed themselves for a prompt not producing the results they expected, despite working with the experimenter to craft prompts. This might limit the creativity and freedom a user should feel when interacting with chat-based AI.
    \item After a few prompts, a participant noted they would start a spreadsheet to maintain the data they were receiving from the model, but thought it would be difficult to switch between the spreadsheet and further prompting. Getting the model to generate a table to help the participant visualize a future spreadsheet was helpful in this situation.
    \item A participant was unsure the model considered the entire context of the prompt it was given, even when this context was in the prompt, and felt there was no way to verify this with the model.
    \item Upon noticing a result of potentially summarized or hallucinated data was given by the model, a participant noted that if they could not trust the results and had to manually search to verify the data in the table. They said they felt it severely limited the benefits of \BingChat.
\end{itemize}

We believe this protocol is an improved adaptation of the traditional Wizard-of-Oz approach for studies of generative AI, since participants interact with a real AI model, but the implementation costs were extremely low.

% we are still able to get these insights as at low of a cost as Wizard-of-Oz methods because prompts were developed beforehand and we did not need to build or implement a new user interface.

% \benz{Is there a way to quantify how much the mediation helped in the development of insights from the experiment?  I guess there wasn't a baseline comparison with having no mediation and seeing what happens.} \ian{yes, we would need to do a comparative study against other current WOZ or other techniques to measure improvements. One strategy might be to collect insights from other research about struggles that end-users are having with prompting and speak on how this method addresses them for formative studies that elicit design principles, which is the goal of this initial study}

% \section{Study of end-user data analysis}
% \subsection{Motivation and related work}
% \subsection{Method}
% \subsection{Results}

\section{Discussion and limitations}
% Participatory prompting may have benefits that parallel those provided by other social workflows like pair programming and apprenticeships.
% \ian{not enough evidence yet to make this claim so will delete}

% \benz{This approach raises the question of whether an AI assistant can be substituted for the human mediator in this method.  How much did the mediator help in improving the prompt and how much could the crafting of the actual prompt from the user input have been done by using an LLM.  Okay - I see you thought of this below. }

% AI as the experimenter to assist in participatory prompting? (however this may take more time than the current protocol, but I'm sure research into this exists)
The participatory prompting approach detailed in this paper raises the question of whether some activities carried out by the human experimenter could be supported with AI. One question that remains to be answered is what affordances human prompt strategies have over an AI that is focused in helping the participant best interact with the generative AI.

However, such a protocol might increase the turn time or number of turns taken as the participant has to interact with this new AI prompting assistant while also performing their sensemaking task. Human-driven participatory prompting also allows the experimenter to ask user experience questions and inquire on participant motivations, which can provide valuable insights for researchers but may not be valuable for the actual prompting and might not be asked by an AI assistant.

One clear extension of this protocol is for the experimenter to also draw upon the library of existing AI plugins and recommend useful experiences that help the participant solve their problem. This could be directly compared to the effectiveness of the existing plugin experience where the model chooses which plugin to use, assuming the user has the installed plugins. 

% \benz{One clear extension of this protocol is for the human mediator to also draw upon a library of existing AI plugins that can help the user solve the problem.  This extension could also be directly compared to having the model directly choose which plugin to use, which is how ChatGPT plugins currently work, assuming the user has installed appropriate plugins.}

% Can't exactly see where user prompting would fail since assisted by experienced experimenters. 
One limitation of this protocol is that because the experimenters will take a turn at helping craft prompts with the participant, results following this protocol may not give a clear understanding of where and when a user's unsupported prompting would have had issues. We attempt to address this limitation by having the participant reflect on how they would have modified a prompt (see Appendix \ref{apx:script}).

% Predisposing spreadsheet/table representations, not as organic on when a user might have this idea.
Similarly, because we are interested in how users might perform sensemaking in spreadsheets by leveraging AI, we have created an environment and crafted prompts that emphasize the use of spreadsheets and organized data. This might mean that the choice to move from prompting to spreadsheeting may not be as organic as it would if the user interacted with the model without experimenter assistance. There is a multitude of data analysis experiences that might also be useful for users (e.g., OpenAI's Code Interpreter, which performs data analysis tasks with Python code \cite{openai_chatgpt_plugins_code_interpreter}). The freedom to choose from available interactions would provide useful insights for user needs for end-user data analysis and sensemaking tasks.

% \benz{Things have moved quickly and the ChatGPT plugin Code Interpreter, which isn't available publicly but will be soon, takes the user into a Python development experience when you have queries related to data-centric tasks.}

% Another issue with the table the model presented was that it dropped data like links to photographs of the places she was interested in, and requested in a previous prompt, once the data being returned was converted into a table. This might be a limitation of the table view not being able to show links as the normal chat view can.

% covered below: At this point, the table the model was generating would stop mid-generation. This worried the participant that further prompting would just further degrade the results. This was seen as a good time to then abandon the current prompting, switch to placing the data in a spreadsheet, and then begin a new prompting session for further exploration.

Some interactions were limited by the inability to re-trigger generation of a response with respect to a specific query within the conversation in \BingChat (re-generation and editing a query is possible for instance, within ChatGPT and the OpenAI playground; this enables a kind of flexibility and fluidity that is akin to being able to independently edit and run different code cells out-of-order in a Jupyter notebook). For instance, if the participant changed their mind about a query, or if the system stopped generating text partway through a response, which worried one participant about continuing to prompt the model. In \BingChat the only option is to submit a follow-up query within the same conversation, but which will include the undesired or incomplete previous queries and responses as part of the ``context''. The alternative is to start a fresh conversation and then laboriously ``replay'' the conversation, building up the same conversational state (or more likely, a \emph{similar} state, since the system's responses are nondeterministic) through the same series of prompts until you arrive at the point in the conversation at which you wish to try a different query. Neither of these options is practical or predictable in a time-limited study.

\section{Conclusion}

In this work-in-progress paper we have presented the ongoing development of \textbf{participatory prompting}: a lightweight user research method for eliciting opportunities for AI assistance in knowledge workflows. It uses a real, functional generative AI system, thus improving on traditional Wizard-of-Oz or paper prototyping, where the user interaction can become unmoored from the technical reality of these systems. On the other hand, it allows researchers to use an ``off-the-shelf'' AI model with no additional engineering costs for fine-tuning, customization, or UI development, enabling rapid and broad-ranging testing of user experiences.

We reported a pilot study (n=2) in which we tested the participatory prompting protocol. The pilots have resulted in improvements to the protocol, changes to the script, and reflections on how to get the most insight out of a participatory prompting session. The pilots have validated the feasibility of the protocol as a method for understanding the user experience of generative AI in knowledge workflows. In future work, we are planning to run a full-scale participatory prompting study to elicit opportunities for AI assistance in the data analysis workflows of end-user programmers in spreadsheets.

% \section{Acknowledgements}

\bibliography{references}
\bibliographystyle{apacite} 

\appendix
\newpage
\section{Study script}
\label{apx:script}
\subsection{Materials/activities pre-interview}
Ask to prepare spreadsheet / reflect on data-driven workflows.

\subsection{[5 minutes] Opening}
Introductions and pleasantries, consent form, demographics form.

\subsection{[10 minutes] Discussion of current data decision practices.}
\begin{itemize}
    \item Can you briefly describe your role?
    \item Can you describe, with examples, what kinds of data-driven decision making you do as part of your role?
    \item Can you describe, with examples, what tools you use?
    \item Can you describe, with examples, how you approach an unfamiliar data-driven decision making problem? An unfamiliar problem where you had to make a decision based on some data. This could be tabular data, lists, or text, etc.
    \item Can you describe an unfamiliar data decision problem, potentially fictional, you may encounter in the future?
\end{itemize}

If this produces a satisfactory scenario, proceed to turn taking, else ask:
Can you share an example of a problem that required you to develop a new workflow?

\subsection{[30 minutes] Participatory prompting, turn taking}

Per turn:
\begin{itemize}
    \item Is this useful? Why or why not?
    \item Are you confused, surprised, or indirectly inspired?
\end{itemize}

To progress, choose one or more of:
\begin{itemize}
    \item What would you like to know next? What else would you need to know to follow these suggestions?
    \item What would you change in your query to make this more useful? Would you ask this a different way?
    \item (Experimenter driven, at most once in a row) Let’s see what happens if we try (X). Alternatively: I propose to continue by X, what would you have done (e.g., continued by Y, different task, abandon tool)?
    \item (If issue with result) I see that there is an issue here with (X), if we do this (new prompt) we can get that data back for you (if participant wants this, continue, else 1st question).
    \item (If participant is stuck, only thinking in terms of ``classical'' search engines) Imagine you’re talking to a colleague, and bouncing ideas off them.
\end{itemize}

\subsection{[15 minutes] Post activity interview}
How would a tool like this fit or not fit into your workflow?
If the participant says something like ``If it could do X that would be helpful'', try it out, get feedback from the participant, but keep time in mind.

\begin{enumerate}
    \item What benefits does this hybrid spreadsheet-chat workflow provide over your existing workflow?
    \item When you were surprised/inspired by X (from turn taking), what features/capabilities would be useful in exploring this inspiration further?
    \item What features do you believe would increase the frequency and effectiveness of these inspiring results/moments (e.g., visualizations, videos, suggested prompts)?
    \item How do you audit data/decisions now and how would that change with these AI-powered features?
    \item How would your decision making workflow change with a tool like this?
    % \item Should this be part of Excel, or a separate tool?
    % \item If Excel, what does putting AI in Excel give you over putting Excel in the separate tool?
    % \item If separate tool, what features of Excel might be required to make these external spreadsheets as effective as when they are located within Excel?
    \item What barriers or frustrations did you have with this experience that prevented you from exploring the question to your satisfaction?
    \item What are the advantages or disadvantages of using a chat-based interface?
\end{enumerate}

\newpage
\section{Pre-identified prompts}
\label{apx:prompts}

This section lists prompts that we have determined through trial and error for use during the study.

\begin{enumerate}
	\item Problem conceptualization, decomposition, identifying parts of the problem that could be tackled in a spreadsheet
		\begin{enumerate}
			\item <Description of user problem>. Explain how to use a spreadsheet for this with an example.
			\item Explain a different way to use a spreadsheet for this with an example.
			\item I am trying to make a data-driven decision about <X>. Is this a good problem to use data tools such as spreadsheets to solve? Explain why or why not. What sub-problems or related problems are good candidates for spreadsheet solutions? Justify your answer.
		\end{enumerate}
	\item Identifying relevant datasets
		\begin{enumerate}
			\item What data is relevant to this problem. List sources.
			\item Use an online data source to add a useful column to this table. Prove your sources are real.
			\item Add a column to the table containing a score representing <X>. Invent a criterion for this score based on publicly available information. Justify your criterion.
			\item Add more rows and columns to the table based on information you consider relevant to the decision of <X>.
			\item Add columns to the table with data from the web such as <X>. Cite your sources.
			\item Use information from the web to populate the spreadsheet with more accurate figures. Cite your sources.
			\item Make an example spreadsheet according to your suggestions above. Use information from the web to populate the spreadsheet with accurate information. Cite your sources.
		\end{enumerate}
	\item Figuring out how to clean and structure data
		\begin{enumerate}
			\item Explain how to put this data in a spreadsheet with an example.
		\end{enumerate}
	\item Developing an analytical strategy, involving application of multiple features in multiple steps
		\begin{enumerate}
			\item Explain how to <user problem> in Excel with steps.
			\item Explain another way to <user problem> in Excel with steps.
			\item It is not possible to <suggestion>. Explain an alternative method with steps.
			\item What spreadsheet features can I use, such as charts, formulas, conditional formatting, pivot tables, etc. Show examples.
			\item <Model suggestion> Explain how to do this with an example.
		\end{enumerate}
	\item Learning how to use relevant features
		\begin{enumerate}
			\item Explain how to use <feature> to solve this problem in Excel with an example.
		\end{enumerate}
	\item Exploration of alternative analyses
	\item Presenting and communicating results
\end{enumerate}
Others (non-categorised)
\begin{itemize}
	\item Make a spreadsheet example
	\item <Model mistake> is not correct. Provide an alternative answer and prove that your answer is correct.
        \item Sometimes \BingChat will refuse to make a spreadsheet. Try asking for a `table' instead. Or ask repeatedly.
\end{itemize}

\end{document}